\documentclass[12pt,onecolumn,twoside]{IEEEtran}

\usepackage{listings}
\usepackage{amsmath}
\usepackage{graphicx}
\usepackage{amsfonts}
\usepackage{array}
\usepackage{amssymb}
\usepackage{helvet}
\usepackage{color}
\usepackage{rawfonts,amsfonts,psfrag,flushend,geometry}
\usepackage{bm}
\usepackage{float}
\usepackage{multirow}
\usepackage{cite}
\usepackage{algorithmic}
\usepackage{textcomp}
\usepackage{xcolor}
\usepackage{verbatim}
\usepackage{bm}
\usepackage{tabularx}
\usepackage{colortbl,booktabs}
\usepackage{subfigure}
\usepackage{mathrsfs}
\usepackage{ulem}
\usepackage{caption}

\begin{document}
\bibliographystyle{IEEEtran}
\title{FusionNet: Enhanced Beam Prediction for mmWave Communications Using  Sub-6GHz Channel and A Few Pilots}
\author{Chenghong Bian, Yuwen Yang, Feifei Gao, and Geoffrey Ye Li
\thanks{C. Bian,  Y. Yang, and F. Gao are with Institute for Artificial Intelligence, Tsinghua University (THUAI),
Beijing National Research Center for Information Science and Technology (BNRist),
Department of Automation, Tsinghua University, Beijing, P.R. China,  100084, P.R. China (email: \{bianch16,yyw18\}@mails.tsinghua.edu.cn, feifeigao@ieee.org).}
\thanks{G. Y. Li is with the School of Electrical and Computer Engineering, Georgia Institute of Technology, Atlanta, GA, USA (email: liye@ece.gatech.edu).}
}
\maketitle \thispagestyle{empty}

\vspace{-15mm}

\begin{abstract}
In this paper, we propose a new downlink beamforming strategy for mmWave communications using uplink sub-6GHz channel information and a very few  mmWave pilots. Specifically, we design a novel  dual-input neural network, called FusionNet, to extract and exploit the features from sub-6GHz channel and a few mmWave pilots to accurately predict mmWave beam.  To further improve the beamforming  performance and avoid over-fitting, we develop two data pre-processing approaches utilizing channel sparsity and data augmentation. The simulation results demonstrate superior performance and robustness of the proposed strategy compared to the existing one that purely relies on the sub-6GHz information, especially in the low signal-to-noise ratio (SNR) regions.
\end{abstract}


\newpage

\section{Introduction}

Millimeter-Wave communications over 30--300GHz band offer a large transmission bandwidth and has been deemed as the key compensation for the current sub-6GHz wireless spectrum\cite{6515173,bai2014coverage}. To combat the severe path loss in mmWave band, the transmitter at the base station (BS) generally deploys a large number of antennas and forms a highly directional beam towards users \cite{6600706,7010533}, which requires accurate downlink channel state information (CSI) \cite{7400949}. However, the increasing in the number of antennas requires significant training overhead to obtain accurate downlink CSI.

Many techniques have been proposed to utilize the channel sparsity to reduce the training overhead of the mmWave downlink transmission \cite{7178503,6847111,6555020,7511578,8917662}. For instance, Compressive sensing (CS) channel estimation for mulit-user massive MIMO systems in \cite{7178503} has investigated the number of measurements required for reasonable performance. The angle sparsity has been observed in  \cite{7511578} and used to develop a structured CS based channel estimation scheme. Moreover, the CS based algorithm in \cite{8917662} exploits the block-sparsity nature of mmWave channels in the frequency domain.

Since future wireless communication systems are expected to employ different frequency bands, it is possible to utilize the CSI features of another frequency band to assist the transmission of the current band. Out-of-band (OOB)\footnote{Actually,  OOB information is not limited to the channel estimated at different frequency band, but would possibly contain a broad categories, including the channels estimated at a different positions \cite{yyw2019deep}, the coordinates of user's position, the radar echo received at the BS, or even the visions captured by the BS camera.  Details can be found in  \cite{8198818} and will not be further expanded here.}  information has been used to reduce the training overhead in many recent works \cite{8198818,7888146,7218630,7413982,8114345}. It has been shown in \cite{7888146}  that 90$\%$ percent of paths is with common angles at frequencies far apart ranging from 900MHz to 90GHz.  In \cite{7218630}, overhead-free multi-Gbps mmWave communication has been established with the out-of-band direction inference obtained from sub-6GHz band. From \cite{7413982}, the downlink covariance is inferred from the observed uplink covariance. In \cite{8114345}, spatial information has been extracted from sub-6GHz channels for beam selection in mmWave band.
The work in \cite{8114345} has opened a new door for beam prediction. However, more work needs to be done to address more complicated channel environments in practical scenarios.

Recently, deep learning (DL) \cite{article} has been applied to a large variety of problems in wireless communications for modulation recognition\cite{8054694}, channel estimation \cite{8640815,8353153}, signal detection \cite{article2,8227772}, channel equalization\cite{943148}, CSI feedback\cite{8322184,8972904}, and end-to-end transceiver\cite{8663966,8645416,8985539}. A deep nerual network (DNN) can approximate any unknown or nonlinear relationship by learning from data, which makes it possible to perform beam selection for mmWave transmission from OOB information. In \cite{8445969,8503086}, DL-based mmWave beam selection in a V2I scenario has been investigated where the size and position of the car serve as OOB information. The framework in \cite{9129762} selects mmWave beam with the help of 3D scene data. Moreover, The DNN designed in \cite{alrabeiah2019deep,9034044} directly obtains the optimal beam given the channel state information (CSI) of the sub-6GHz channel. Similarly, a convolutional neural network (CNN), with fewer number of parameters, in \cite{9128825} leverages the sub-6GHz CSI to find the optimal mmWave beam. With the assist of various OOB information, all above approaches greatly reduce the training overhead for the mmWave downlink transmission. However, these beam prediction approaches directly treat the neural network as a black box \cite{alrabeiah2019deep,9034044}, which prevents from further performance improvement, especially when the signal-to-noise ratio (SNR) of the sub-6GHz channel is low.

In this paper, we develop a novel dual-input neural network, called FusionNet, to predict the optimal beam using both the sub-6GHz channels and a very few pilots in mmWave band. Even if with only a few pilots, this strategy can effectively tune the deviated beams to the correct direction and significantly improve the performance compared to its black box counterparts.
The FusionNet also enables to exploit the channel sparsity in the angular-delay domain for further performance improvement. Moreover, a novel data augmentation approach is also developed to alleviate the over-fitting issue of FusionNet. Numerical results manifest that the proposed FusionNet outperforms the existing strategies in terms of both the prediction accuracy and achievable rate, especially at low SNR regions.

The rest of the paper is organized as follows. In Section II, we present the system model and channel estimation at dual frequency bands. In Section III, we briefly introduce the existing beam prediction using sub-6GHz channel only. Section IV provides the architecture of the proposed neural network and analyzes its complexity. The dataset generation, data transformation, and data augmentation are described in Section V. Numerical results are provided in Section VI, followed by the conclusion in Section VII.

Throughout our discussions, the  scalar variables are represented with normal-face letters $x$ while matrices and vectors with \textbf{upper} and \textbf{lower} case letters, $\mathbf{x}$ and $\mathbf{X}$, respectively. Transpose and Hermitian operators are denoted by $(\cdot)^T$, $(\cdot)^H$, respectively. The $l_2$ norm is denoted as $||\cdot||_2$ and $|{\cal C}|$ is the cardinality of set ${\cal C}$; $j = \sqrt{-1}$ is the imaginary unit and $(\mathbf{a})_c$ denotes the $c$th entry of vector $\mathbf{a}$.
\begin{figure}[t]
\centering
\subfigure[The BS and the mobile user communicate over both sub-6GHz and mmWave bands with co-located sub-6GHz and mmWave antennas.]{
\includegraphics[width=8cm,height=4cm]{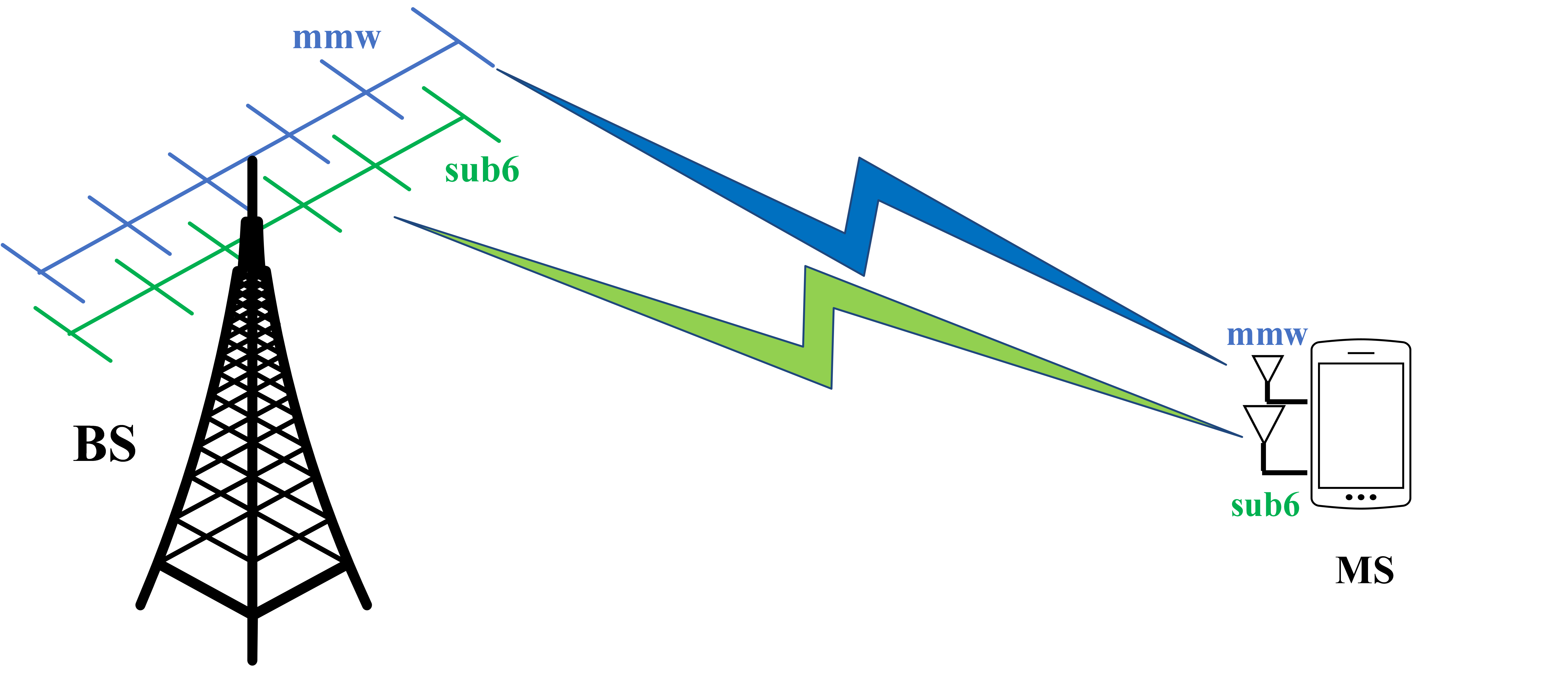}
}
\vspace{5pt}
\subfigure[The mmWave band transmission contains only one single RF chain that connects to the mmWave antennas via phase shifters and switchers.]{
\includegraphics[width=8cm,height=4cm]{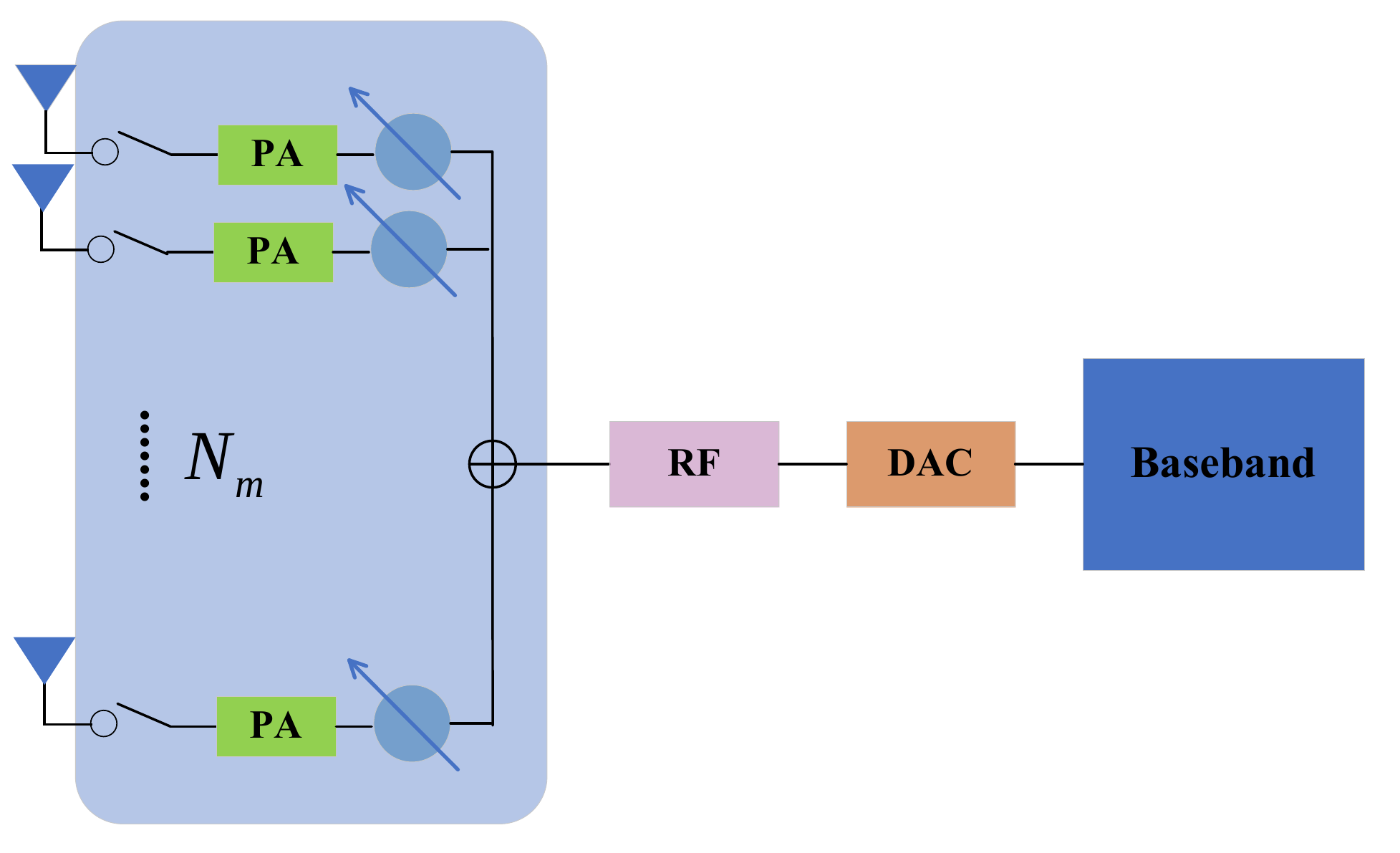}
}
\captionsetup{font={small}}
\caption{\small  The communication scenario and the mmWave architecture. }
\label{fig1}
\end{figure}
\section{System Model}

Consider a communications system over both sub-6GHz and mmWave bands, which has one BS and one user, as illustrated in Fig. \ref{fig1}. The BS has two sets of antenna arrays, one with $N_{s}$ antennas for sub-6GHz band while the other with $N_{m}$ antennas for mmWave band. The sub-6GHz antenna array is fully digital, where each antenna connects to an independent RF chain. For simplicity, we assume the mmWave antenna array is analog, i.e.,  all antennas connect to a single RF chain via $N_{m}$ phase shifters and $N_{m}$ switchers \cite{7400949}, as shown in Fig. \ref{fig1}.  The user has two antennas, working at sub-6GHz and mmWave bands. Both sub-6GHz and mmWave communication links use orthogonal frequency division multiplexing (OFDM) with $K_s$ and $K_m$ subcarriers, respectively. The codebook of the downlink beamforming at mmWave band is denoted as ${\cal C}=\{\mathbf{f}_1, \mathbf{f}_2, \ldots, \mathbf{f}_{|{\cal C}|}\}$. The target here is to find the optimal downlink beamforming index at the mmWave band via  the estimated uplink channel at sub-6GHz and a few mmWave pilots.

\subsection{Uplink Channel Estimation at Sub-6GHz}

Let $\mathbf{h}_{s}[k] \in \mathbb{C}^{N_{s}\times1}$ denote the uplink channel vector of the $k$-th subcarrier at the sub-6GHz band. The received uplink signal at the BS can be expressed as
\begin{equation}
    \mathbf{y}_{s}[k] = \mathbf{h}_{s}[k]s_{s}[k] + \mathbf{n}_{s}[k]
\end{equation}
for $k = 1,2,\cdots,K_s$, where $s_{s}[k]$ is the pilot of the $k$th subcarrier and $\mathbf{n}_{s}[k]\in \mathcal N (0,\sigma^2)$ is the corresponding noise.
Either least-square (LS) or linear minimum mean-squared error (LMMSE) channel estimation, in the frequency domain or the time domain can be used, the detailed process is omitted here.

\subsection{Partial  Channel Estimation at MmWave Band}

Since there is only one RF chain, the regular uplink or downlink channel estimation at mmWave band needs to be repeated $N_{m}$ times via switching the RF chain onto different antennas, or vary the weights of the phase shifters when each single RF chain connects to $N_{m}$ antennas simultaneously. Such process costs a large amount of time resource, especially when $N_m$ is large.

Here, we estimate the channels only on $\tilde{N}_m \ll N_{m}$ antennas to assist the beam prediction from the sub-6GHz to the mmWave band.  The mmWave antennas that participate in the mmWave uplink channel estimation is called as {\it active antennas}. Note that the estimated channels corresponding to this $\tilde{N}_m$ antennas may not even be sufficient to recover the whole $N_{m}$ channel elements via compressive sensing or the deep learning techniques. To make the overall illustration clear, we will first present how to estimate the uplink $\tilde{N}_m$ channels by changing the weights of the phase shifters.\footnote{The other  way that switches the RF chain to the $\tilde{N}_m$ antennas sequently is not stable in practical applications and will not be adopted here.}

The switchers of $(N_m-\tilde{N}_m)$ antennas are in off status during the training stage and the mmWave RF chain connects to the $\tilde{N}_m$ antennas via the $\tilde{N}_m$ phase shifters, simultaneously. Denote the channel on the $k$th subcarrier of these $\tilde{N}_m$ antennas as $\tilde{\mathbf{h}}_{m}[k]$. In order to estimate these  $\tilde{N}_m$ channels, the user should send the training OFDM block $\tilde{N}_m$ times since there is only one RF chain. The pilot signal on the $k$th subcarrer in the $i$th training block is denoted as $s_{m,i}[k], i=1, \ldots, \tilde{N}_m$. For simplicity, we assume that the pilot on all $K_m$ subcarriers and all $\tilde{N}_m$ training blocks are same i.e., $s_{m,i}[k]=1,\forall(i,k)$.\footnote{It is also possible to consider the combo type training where only part of the subcarriers are used for pilots while the rest are used for unknown data. }

Moreover, let us denote $\tilde{f}_{i,j}$ as the value of the phase shifter for the $i$th training block and the $j$th antenna, $i,j=1, \ldots, \tilde{N}_m$, which is universal for all subcarriers. Denote $\tilde{\mathbf{f}}_i=[\tilde{f}_{i,1},\tilde{f}_{i,2},\tilde{f}_{i,\tilde{N}_m}]^T$  with $|\tilde{f}_{ij}|=1$. As in Fig \ref{fig1}.b, the received  signal on the $k$th subcarrier of the $i$th training block is
\begin{align}
y_{i}[k]=\tilde{\mathbf{f}}_i^T\tilde{\mathbf{h}}_{m}[k]+n_{i}[k],\quad  k=1,\ldots, K_m.
\end{align}
Stacking $y_{i}[k]$ from $\tilde{N}_m$ training block into one vector, we have
\begin{align}
\mathbf{y}[k]&=[y_{1}[k],y_{1}[k],\ldots,y_{\tilde{N}_m}[k]]^T=\tilde{\mathbf{F}}\mathbf{h}_{m}[k]+\mathbf{n}[k],
\end{align}
where $\tilde{\mathbf{F}}=[\tilde{\mathbf{f}}_1, \tilde{\mathbf{f}}_2, \ldots, \tilde{\mathbf{f}}_{\tilde{N}_m}]^T $ and $\mathbf{n}[k]=[n_{1}[k], n_{2}[k],\ldots, n_{\tilde{N}_m}[k]]^T.$ Normally, $\tilde{\mathbf{F}}$ is selected as the $\tilde{N}_m\times \tilde{N}_m$ normalized  DFT matrix during the training process.

\subsection{Downlink Data Transmission at mmWave Band}

During the downlink data transmission, the BS uses all $N_{m}$ antennas to form one single narrow beam, $\mathbf{f} \in {\cal C}$. Since we assume only one RF chain, the downlink
beamforming can be represented as $ \mathbf{f}=[f_1, f_2,\ldots, f_{N_{m}}]^T$ with $|f_i|=1$. The received  signal on the $k$-th subcarrier in one OFDM block can be written as
\begin{equation}
    y_{m}[k] = \mathbf{h}^T_{m}[k] \mathbf{f} s_d[k] + n_{m}[k],
\end{equation}
where $n_{m}[k]$ and $s_d[k]$ represent the corresponding noise and signal with powers $\sigma_n^2$ and $P_s$, respectively.

The achievable rate of the downlink transmission can be expressed as
\begin{equation}
    R(\mathbf{h}_{m},\mathbf{f}) = \sum_{k=1}^{K}\log\left(1+\frac{|\mathbf{h}^T_{m}[k]\mathbf{f}|^2P_s}{\sigma_n^2}\right).
\end{equation}
Since the phase shifters are generally constrained with limited bits, the size of ${\cal C}$ is finite.  In the previous works \cite{6134486,6600706,5262295}, exhaustive search is performed to find optimal $\mathbf{f}^\star$ to maximize the achievable rate, $R(\mathbf{h_{m}},\mathbf{f})$, according to the channel conditions.

\begin{figure}[t]
\centering
\subfigure[The measured PDP of 3.5 GHz with 100MHz bandwidth in corridor scenario.]{
\includegraphics[width=10.5cm,height=5.3cm]{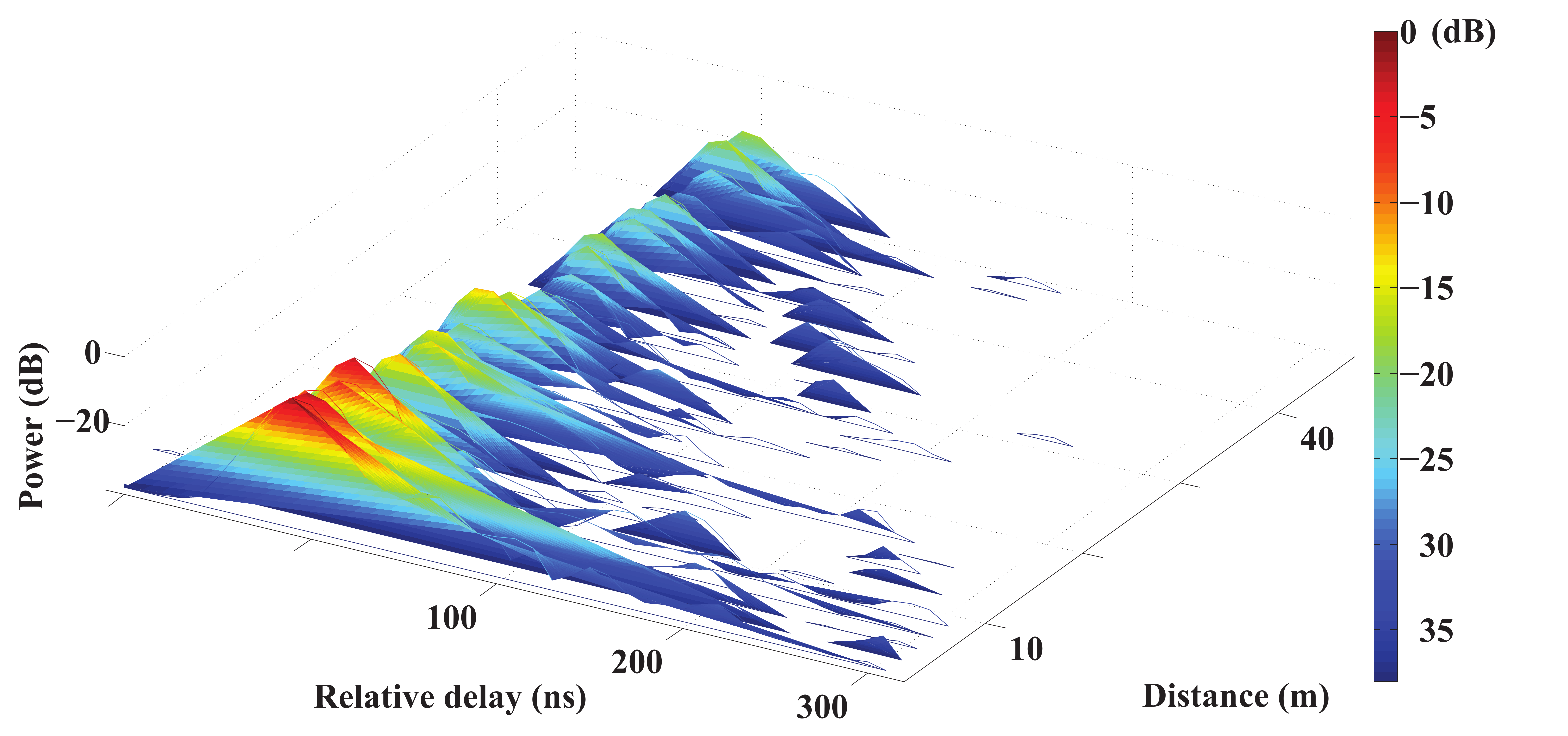}
}
\subfigure[The measured PDP of 28 GHz with 100MHz bandwidth in corridor scenario.]{
\includegraphics[width=10cm,height=5cm]{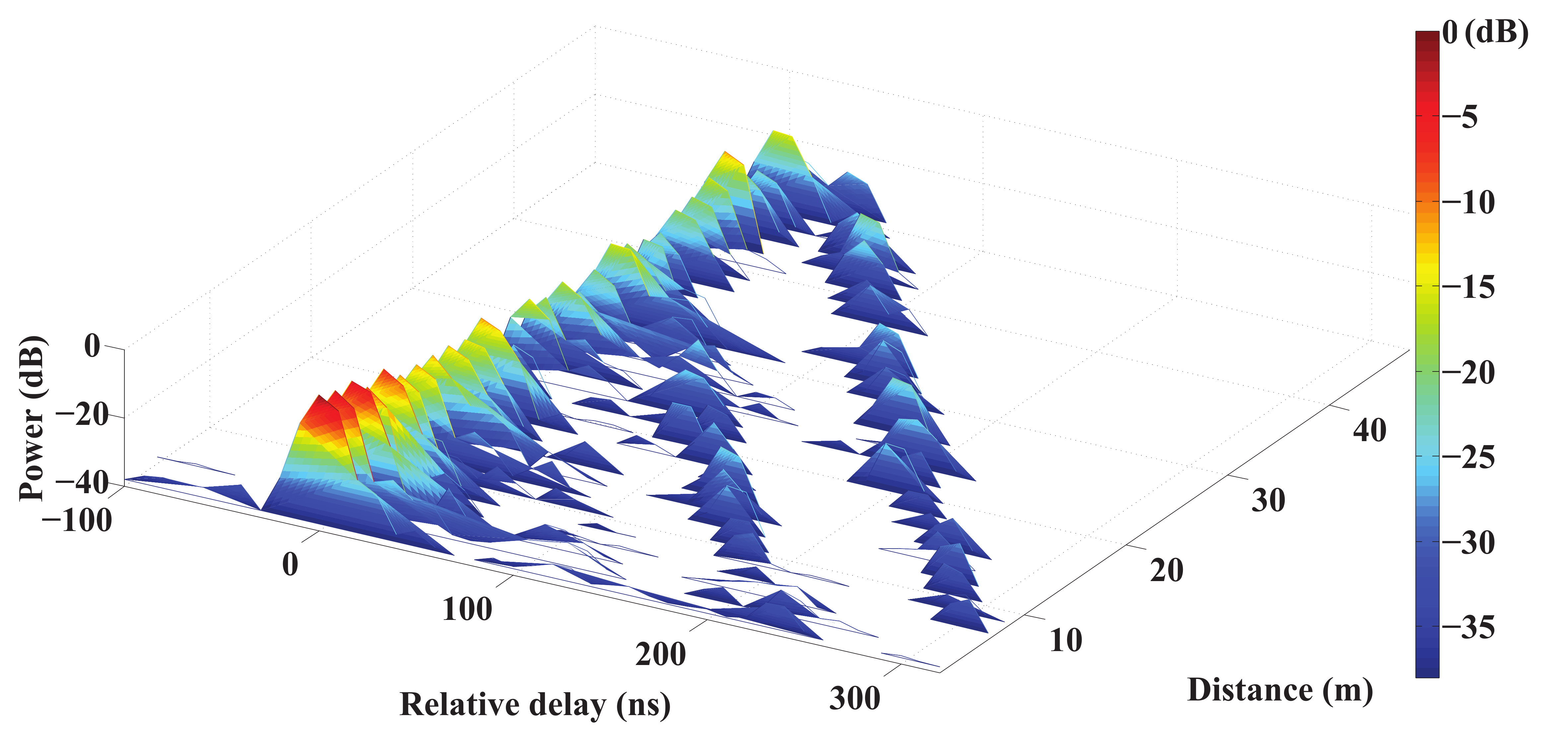}
}
\captionsetup{font={small}}
\caption{\small  The illustration of certain congruency between the sub-6GHz channel and the mmWave channel in the same propagation enviroment.}
\label{fig_congrugency}
\end{figure}

\subsection{Direct Prediction from sub-6GHz Channel}

\textcolor[rgb]{0.00,0.00,0.00}{A typical example is shown in Fig. \ref{fig_congrugency}, where the power delay profile (PDP) of the channels from both 28GHz and 3.5GHz bands with 100MHz bandwidth are measured in the corridor environment\cite{8946574}. From the figure, though the PDP of 28GHz channel is more sparse compared to that of 3.5GHz, the contour of both frequency bands are similar, which inspires to use the congrugency of the channel vectors in \cite{7888146,7413982,8114345}.}  Specifically, it is possible to predict optimal mmWave downlink beam $\mathbf{f}^\star$ directly from the sub-6GHz uplink channel in \cite{alrabeiah2019deep,9034044}. For instance, it has been proven in \cite{alrabeiah2019deep} that there exists a deterministic mapping from the sub-6GHz channel to the mmWave beams. In \cite{9034044}, the optimal mmWave beam is predicted by estimating the PDP of the
sub-6GHz channel, where the PDP is considered as a fingerprint for the UE position and thus contains essential angular information for beam selection in a cell-specific manner with given environments. The training dataset of the two frequency bands is generated from the Wireless System Engineering (WiSE). Then, a DNN similar to  \cite{alrabeiah2019deep} is trained to predict the best beam within the candidate set ${\cal C}$.

Although predicting the mmWave beam from sub-6GHz channel CSI is theoretically and experimentally demonstrated effective. From \cite{alrabeiah2019deep,9034044}, there is still a big gap to improve its performance, especially when the SNR of the sub-6GHz channel is low. In fact, even without channel noise, the prediction accuracy is still below 85$\%$ and does not meet the practical communications requirement, which motivates us to improve the accuracy by combining features of the sub-6GHz channel and pilots and the mmwave channel.

\section{FusionNet for Improved Beam Prediction}
In this section, we will show how to merge a very few mmWave pilots with the sub-6GHz CSI to significantly improve the accuracy of mmWave beam prediction.

\begin{figure}[t]
\centering
\includegraphics[width=13cm]{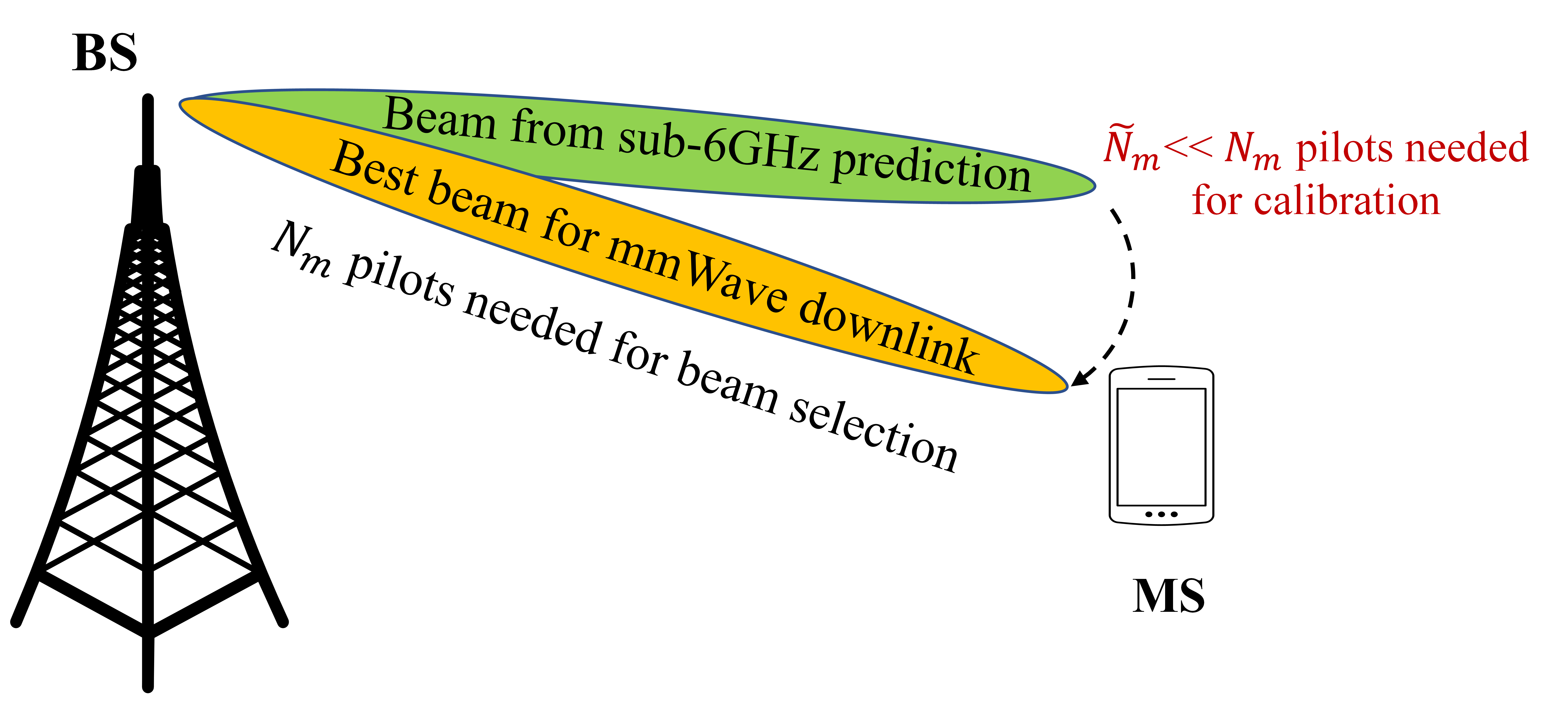}
\captionsetup{font={small}}
\caption{\small Beam selection directly from mmWave antennas requires $N_m$ training resources. As the prediction from sub-6GHz channel already roughly points to the correct direction, we will only need $\tilde{N}_m  \ll {N}_m$ training resources to calibrate and get a better beam prediction.
}
\label{fused}
\end{figure}

\subsection{Why A Few MmWave Pilots?}

The best beam for mmWave downlink almost corresponds to the strongest path in the angular domain. Hence, the beam prediction problem is readily solved if the angular information of the mmWave channel is available. As discussed before, predicting directly from the sub-6GHz channel is not accurate enough due to the following factors:
\begin{itemize}
  \item [1)]
  With limited number of sub-6GHz antennas the BS cannot offer high resolution in the angular domain;
  \item [2)]
  As pointed out in \cite{9064361}, the support for the mmWave channel in the angular domain is only a subset of that for the sub-6GHz channel with the same spatial grid quantization, and thus the angle of the strongest path for sub-6GHz channel may not be the mmWave counterpart;
  \item [3)]
  The DL approximation of the channel mapping from sub-6GHz to mmWave is not accurate enough due to limited number of training samples and limited network size.
\end{itemize}

Though the beam direction predicted from the sub-6GHz channel usually deviates from the true one as we can imagine, it still preserves certain channel spatial information and could serve as a valid starting point to find the best beam. Hence, the limited mmWave pilots would be very helpful to ``calibrate" such deviation and significantly enhance the prediction accuracy, as illustrated in Fig. \ref{fused}. Note that since the mmWave pilots are mainly used for calibration but not estimation, the number of required pilots is much smaller than that to estimate the complete mmWave channel, i.e., $\tilde{N}_m  \ll N_m$.

This key rationale is simple but is  rather practical and useful for beam prediction, channel estimation, or data detection, etc. In all, one should always count on a few pilots to achieve the ultimate precision after using the DL to get a coarse starting point.
In the following,  we will present a detail design of the neural network architecture to effectively merge the mmWave pilots and sub-6GHz channel information, to accurately predict the beam in the mmWave band.

\subsection{Network Architecture}

Following \cite{alrabeiah2019deep},  we here adopt multi-layer perceptron (MLP) to represent the  mapping function from the sub-6GHz channel to the optimal beam $\mathbf{f^\star}\in {\cal C}$.

\begin{figure}[t]
\centering
\subfigure[Directly concatenate two vectors before feeding to the network]{
\includegraphics[width=10.5cm,height=2.7cm]{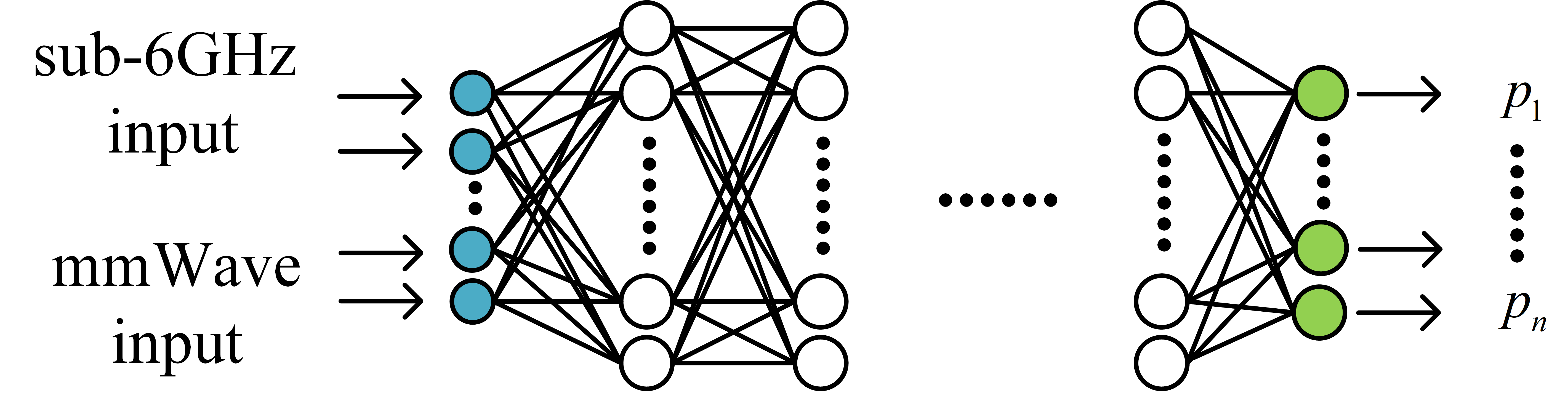}
}
\subfigure[The proposed FusionNet]{
\includegraphics[width=18cm,height=10cm]{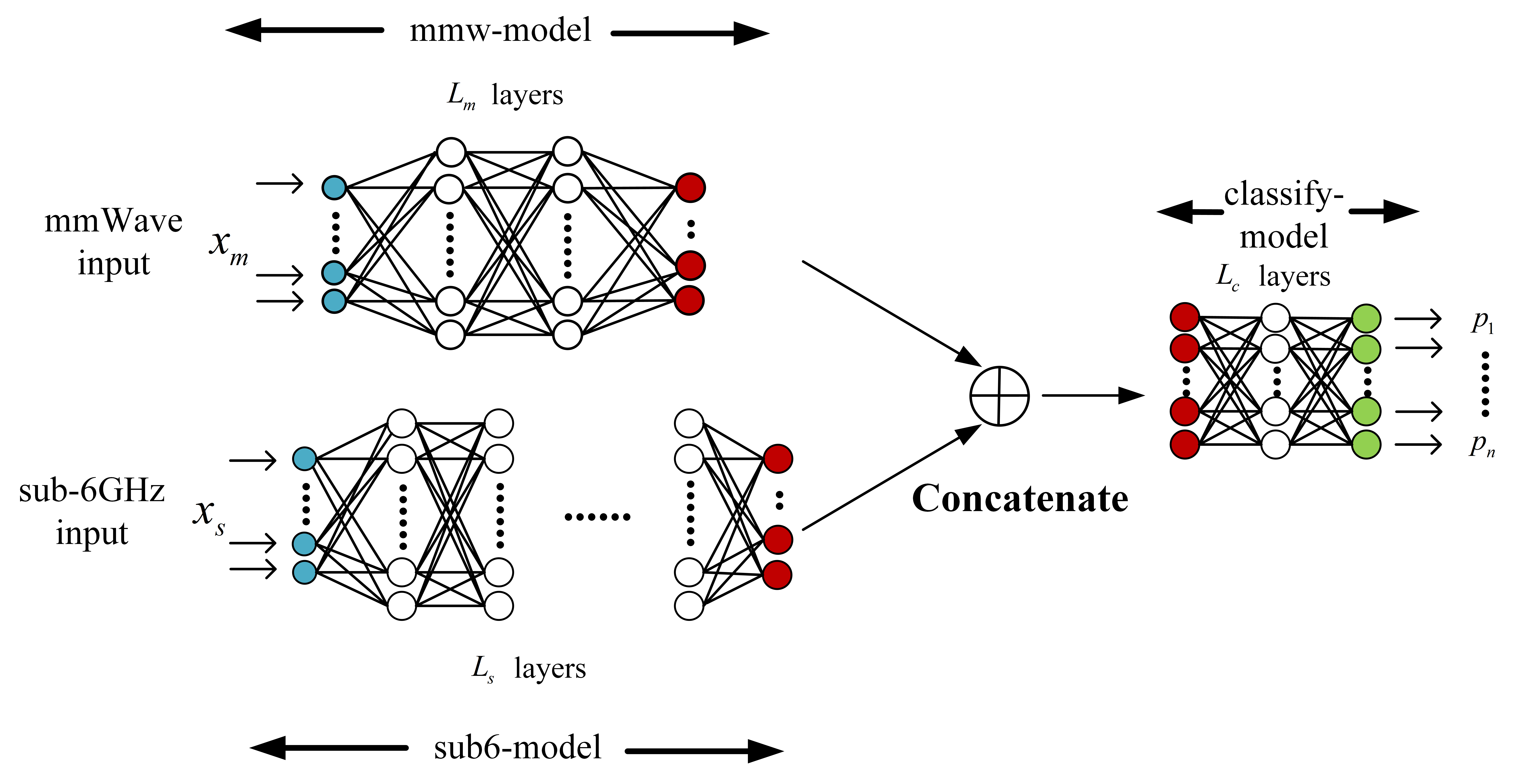}
}
\captionsetup{font={small}}
\caption{\small  The architectures for the shallow model and the proposed FusionNet}
\label{fig2}
\end{figure}
One straightforward way is to directly  concatenate the sub-6GHz channel and the mmWave channel as the input of a DNN, as illustrated in Fig. \ref{fig2}(a), which is known as  the shallow model \cite{multimodal}. However, as the correlation between the sub-6GHz and the mmWave channel is highly non-linear, it is hard for a neural network to learn their individual features effectively from a concatenated vector. The shallow model still has some performance loss, especially when the sub-6GHz channel is with low SNR.

Inspired by \cite{7925709,multimodal}, we design a dual-input network, called FusionNet, as shown in Fig. \ref{fig2}(b). The FusionNet first extracts the features from the sub-6GHz and mmWave channels separately, and fuses them in a concatenation layer to generate a probability vector $\mathbf{p}$ whose $i$th entry, $p_i$, represents the probability for the $i$th beam (in the given codebook $\mathbf{{\cal C}}$) being the optimal one. We denote the inputs corresponding to the sub-6GHz and the mmWave channels as $\mathbf{x}_s$ and $\mathbf{x}_m$, respectively, which are the vectorization of channels on all subcarriers estimated in Section II.

As in Fig. \ref{fig2}(b), the FusionNet is comprised of three sub-networks, i.e., mmw-network, sub6-network and classify-network.  The mmw-network, with $L_m$ fully connected layers, extracts the features, such as angular, delay, and path gain information from the mmWave channel input. The sub6-network, with $L_s$ layers, extracts information from the sub-6GHz input. The classify-network takes the concatenated feature as the input, and is followed by a Softmax layer. Each fully connected layer in these sub-networks is followed by a BatchNorm layer, a Relu layer, and a dropout layer.

Since the frequency discrepancy between the sub-6GHz and the mmWave channel is very large, intuitively, using the sub-6GHz channel to predict the best beam in mmWave band will definitely need more layers compared to using the mmWave channel itself. Hence, we set $L_s > L_m$ in the proposed design.  Denote the numbers of neurons in the $l$th layer in the sub6-network, mmw-network, and classify-network as $n_l^m,n_l^s,n_l^c$, respectively. The mmw-network and sub6-network extract the various path information of the corresponding channels, while the classify-network yields probability vector, $\mathbf{p}$, with length $n_l^c=\mathbf{|{\cal C}|}$. The other parameters of the FusionNet will be further discussed in Section V.

\subsection{Training and Evaluation}
In the training stage, a supervised learning approach is adopted, where the training label, denoted as $\mathbf{t}$, is a one-hot vector representing the best beam for the mmWave downlink transmission. The detailed calculation of $\mathbf{t}$ will be presented later in the next section. We adopt the cross-entropy loss as the loss function:
\begin{equation}
    H_p(\mathbf{t}) = {\sum_{c=1}^{\mathbf{|{\cal C}|}}{(\mathbf{t})_c*\log((\mathbf{p})_c)}},
\end{equation}
which is  minimized by the ADAM optimizer.

In the evaluation stage, new sub-6GHz and mmWave channels will be generated. After pre-processing the data and feeding to the trained FusionNet, the optimal beam can be predicted.

\subsection{Complexity Analysis}
For the FusionNet, the total number of floating point operations (FLOPs) can be computed as:
\begin{equation}
    \sum_{l_m=1}^{L_{m}-1}{n_{l-1}^m n_l^m}+\sum_{l_s=1}^{L_{s}-1}{n_{l-1}^s n_l^s}+\sum_{l_c=1}^{L_{c}-1}{n_{l-1}^c n_l^c}.
\end{equation}

Compared with \cite{alrabeiah2019deep} that only needs to process the training data from the sub-6GHz channel, the first two sub-networks of the FusionNet take training data from the two frequency bands, resulting in a little bit higher complexity. For example, when $L_m=4,L_s=6,L_c = 3$ and $n_l^m, n_l^s = 2048, n_l^c=64$, the total complexity is mainly determined by the first two sub-networks and is approximately twice of the complexity in \cite{alrabeiah2019deep}. Nevertheless, the neural network is mainly trained offline and is deployed online at the BS, where the computing resources are always assumed abundant.

\section{Dataset Generation and Data Preprocessing}
In this section, we will first introduce how to generate the training data with the corresponding optimal beam labels. Then two novel data pre-processing methods
will be designed to further improve the performance of the FusionNet.

\subsection{Data Set Generation}
The dataset for the FusionNet comprises of the mmWave channel, the sub-6GHz channel estimated at different user positions, and the corresponding best beam index.
We first generate the  sub-6GHz and mmWave channels 
using the ``O1 scenario"  in DeepMIMO dataset \cite{Alkhateeb2019}. The parameters used in the data generation process are summarized in Table.~\ref{table1}.

The DeepMIMO dataset contains parameters of the $R$ strongest rays for each user and is represented by ($\psi_r,\theta_r,\alpha_r,\tau_r$), where $\psi_r$ denotes the azimuth angle at the BS of the $r$-th path, $\theta_r$ denotes the elevation angle, $\alpha_r$ is the complex gain, and  $\tau_r$ is the delay. The mmWave channel is constructed using a geometric channel model whose channel vector at the $k$-th subcarrier is
\begin{equation}
    \mathbf{h}_{m}[k] =  \sum_{r=1}^R{\alpha_r {e^{-j\frac{2\pi k \tau_r}{T_s}}}\mathbf{a}(\psi_r,\theta_r)},
    \label{equchgen}
\end{equation}
where $\mathbf{a}$ is the steering vector and $T_s$ is the reciprocal of the OFDM subcarrier interval. The sub-6GHz channel is also generated using (\ref{equchgen}) while the DeepMIMO dataset will automatically adjust the parameters according to the environment and frequency band. Therefore, the geometric channel model is capable to capture the physical characteristics of the signal propagation process including the dependence
on the environment geometry, materials, frequency band, etc., which are of vital importance for DL based beam prediction.

We will place pilots on all subcarriers of the OFDM blocks at both frequency bands. With the received uplink signals, the BS will estimate channels $\mathbf{h}_{m}[k]$ and $\mathbf{h}_{s}[k]$ at the two frequency bands as illustrated in Section II. Similar data normalization \cite{alrabeiah2019deep} is then carried out for input signals in both frequency bands.

Let $\mathbf{h}_m^u[k]$ be the mmWave channel vector at the $k$th subcarrier for the $u$th user from DeepMIMO and $(\mathbf{h}_m^u[k])_n$ be the mmWave channel at the $k$th subcarrier on the $n$th antenna. The channel vectors are normalized by a global normalization value $\Omega$, which is the largest absolute value in the whole dataset, i.e.,
\begin{equation}
    \Omega = \max_{k,n, u}|(\mathbf{h}_m^u[k])_n|.
\end{equation}

After normalization, the magnitudes of all channel elements will be between 0 and 1. We then spilt the normalized complex channel into real part and image part, which will then be stacked together. The final mmWave training dataset can be obtained with size  $U_m\times (2\times K_m \times N_m)$ where $U_m$ is the number of total user positions at the mmWave band.

Next, we will present how to get label $\mathbf{t}$ at each user position. The achievable rate for the mmWave downlink channel of the $u$th user with beam $\mathbf{f}_c$  can be computed as
\begin{equation}
  \kern -10pt R(\mathbf{h}_{m}^u,\mathbf{f}_c) = \sum_{k=1}^{K_m}{\log_2(1+\text{SNR}|\mathbf{h}_{m}^u[k]\mathbf{f}_c|^2)}, \nonumber\\
   \label{equrate}
\end{equation}
for $c = 1,2,\cdots,\mathbf{|{\cal C}|,}$ where SNR denotes the signal-to-noise ratio at the transmitter (user's side). Then  the index of the best beam can be obtained from   the offline searching by
\begin{equation}
    (c^{u})^\star = \arg\max\limits_{c = 1,2,\cdots,\mathbf{|{\cal C}|}}R(\mathbf{h}_{m}^u,\mathbf{f}_c).
    \label{equlabel}
\end{equation}
An one-hot vector $\mathbf{t}^u$ can be obtained for each user to represent $(c^{u})^\star$, whose $(c^{u})^\star$th element is 1 while other elements are 0, which serves as the label in the training and validating stage.

\subsection{Utilizing Channel Sparsity}

In Section II, the estimated sub-6GHz and mmWave channels, $\mathbf{h}_s[k]$ and $\mathbf{h}_m[k]$, are in the spatial-frequency domain. By stacking these channel vectors together, channel matrices for the sub-6GHz band and the mmWave band can be obtained  as
\begin{align*}
    &\mathbf{H}_m^{sf} = \left[\mathbf{h}_m[1],\mathbf{h}_m[2],\cdots,\mathbf{h}_m[K_m]\right],\\
    &\mathbf{H}_s^{sf} = \left[\mathbf{h}_s[1],\mathbf{h}_s[2],\cdots,\mathbf{h}_s[K_s]\right].
\end{align*}

\begin{table}[t]
  \centering
  \begin{minipage}[t]{0.8\linewidth} %
  \captionsetup{font={small}}
  \caption[\small Wireless Insite parameters]{\small Parameters  to generate the channel vectors}
  \label{table1}
    \begin{tabularx}{\linewidth}{|X|X|X|}
      \toprule[1.5pt]
      { parameters} & { mmWave} & { Sub6 GHz}\\\midrule[1pt]
      carrier frequency & 28GHz & 3.5GHz \\
      BS antennas & 64 & 4\\
      antenna interval & 0.5 & 0.5\\
      OFDM band width(MHz)    & 0.5 & 0.02\\
      OFDM Subcarriers    & 512 & 32 \\
      Path  & 5 & 15 \\
      \bottomrule[1.5pt]
    \end{tabularx}
  \end{minipage}
\end{table}
Inspired by \cite{9064361,7727995,7174558,8354789}, the FusionNet can improve the prediction performance by leveraging the sparsity of the channel matrices. Since the sparsity at the two frequency bands are similar, we will only discuss the mmWave channel matrix as an example.  A 2-D Discrete Fourier Transform (DFT) is performed on $\mathbf{H}_m^{sf}$ to find the new channel matrix in the angle-delay domain,
\begin{equation}
    \mathbf{H}_m^{ad} = \mathbf{F}_a \mathbf{H}_m^{sf} \mathbf{F}_d^H,
\end{equation}
where $\mathbf{F}_a$ and $\mathbf{F}_d$ are $N_m \times N_m$ and $K_m \times K_m$ normalized DFT matrices.

As illustrated in \cite{8322184}, limited scattering of channels as well as the large number of antennas at the BS (i.e the large $N_m$) ensure the sparsity of $\mathbf{H}_m^{ad}$  in the angular-delay domain, i.e.,  only $N_a \ll N_m$ rows and $N_d \ll K_m$ columns of $\mathbf{H}_m^{ad}$ have significant values. 
Nevertheless, since there are a limited number of sub-6GHz antennas in the considered scenario and only $\tilde{N}_m\ll N_m$ mmWave antennas are used for  channel estimation, the condition that a large number of antennas are employed at the BS is not satisfied. Thus, we will only adopt one dimension DFT to obtain the sparse representation, $\mathbf{H}_m^{sd}$, in the delay domain by
\begin{equation}
    \mathbf{H}_m^{sd} = \mathbf{H}_m^{sf} \mathbf{F}_d^H.
    \label{dft}
\end{equation}
The sparse representation for the sub-6GHz channel can also be leveraged following (\ref{dft}).
Both the sparse channels, $\mathbf{H}_m^{sd}$ and $\mathbf{H}_s^{sd}$, will be used as the input data to train the FusionNet.

\subsection{Data Augmentation}

The proposed FusionNet is mainly comprised of fully connected layers with a large number of parameters. If trained with  insufficient number of data, the neural network will most likely be over-fitted on the training set and would fail to yield a good generalisation on the test set. Many techniques, e.g., the dropout and batch normalization function, have been used to solve the over-fitting problem. However, as the number of parameters increases, the flexibility of the network becomes extremely high and these techniques fall out too.

Data augmentation has been used in DL to generate additional training data and has achieved great success in the realm of speech recognition\cite{9053008}, image classification\cite{7797091,7393462,antoniou2017data} and deep reinforcement learning \cite{reinforce}. Despite various applications in other areas, the data augmentation approach has not yet been used to generate extra data for the wireless communications, to the best of the authors' knowledge. We next introduce a novel data augmentation approach to generate new artificial samples as the input of the FusionNet.

A plausible data augmentation transformation should preserve label information. That is, one can perform any kind of transformation on the mmWave and the sub-6GHz channels as long as the transformed channels yield the same label as before. From the achieved rate objective (\ref{equlabel}), a simple observation  is that if the mmWave channel vector, $\mathbf{h}_m^u[k]$, is  multiplied by any $e^{-j2\pi \phi}$, then both the achievable rate and the corresponding best beam index for each user remain the same.
This observation inspires us to insert a random phase $\phi$ into $\mathbf{h}_{m}$ to augment the mmWave data. We generate $\phi$ from the uniform distribution, i.e. $\phi \sim \mathcal U(0,1)$. Note that, different subcarriers for a  specific user should share the same $\phi$ while different users may have different $\phi$'s.

Data augmentation is also needed for the sub-6GHz channel. At the first look, it seems hard to directly tell what kind of transformation performed on the sub-6GHz channel will not affect the original label because the sub-6GHz channel does not directly determine the optimal beam (training label). Actually, the sub-6GHz channel provides underlying information of the mmWave channel, such as angular feature and frequency feature, about the propagation environment. Therefore, any kind of transformation that preserves the information in sub-6GHz channel is valid. Hence, multiplying a random value to the sub-6GHz channel vector is also plausible since this linear transformation will not cause any loss of channel information. For simplicity, we augment the sub-6GHz data by
\begin{equation}
    \hat{\mathbf{h}}_{s}^u = \mathbf{h}_{s}^u e^{-j2\pi \chi},
\end{equation}
where $\chi$ is a random phase with $\chi \sim \mathcal U(0,1)$.

Using the above data augmentation approaches, a large number of new training samples can be generated. However, as the number of training samples increases, the computational complexity increases accordingly and the computer memory may also become insufficient. Most importantly,  their underlying information  remains the same even  though the number of synthesized samples grows. As a result,   the neural network's performance would not be  further improved when the number of the augment samples reaches a certain value.

\section{Simulation Results}

In simulation, the neural network is trained using the data with labels described in Section V. During the training phase, Pytorch 1.3.0 is adopted as the DL framework running on a server with RTX 2080 Ti GPU. The number of neurons in each fully connected layer of the mmw-network and sub6-network is 2048 while the number of neurons in each fully connected layer of classify-network is 64. A varying learning rate is adopted, changing from the initial value, $10^{-3}$, to $10^{-4}$ after half of the total epochs, and further to $10^{-5}$ for the last $1/10$ epochs. The number of samples is approximately $1.08\times 10^5$, where 70$\%$ for training and $30\%$ for validation.
The batch size is 512 and the total epochs is 60. To relieve the over-fitting problem, dropout layers are added after the fully connected layers with a dropout rate 0.4.  The performance of the proposed neural network is evaluated in terms of  best beam selection accuracy, i.e., top-1 accuracy  and the corresponding achievable data rate (\ref{equrate}). Specifically, the top-1 accuracy $Acc_{top1}$\footnote{Note that in previous many works \cite{alrabeiah2019deep,9034044}, the accuracy of top-3 beamfomer is also adopted as an import criterion for the beam prediction since the $Acc_{top1}$ is low. For our work, thanks to the essential calibration effect of the mmWave pilot, the top-3 accuracy nearly approaches 100 $\%$ even with low SNR  which can be seen in the following simulations.} is defined as
\begin{equation}
    Acc_{top1} = \frac{1}{N_{test}}\sum_{k=1}^{N_{test}}{\mathbb I_{\hat{c}_k=c^\star_k}},
\end{equation}
where $N_{test}$ is the number of testing data, $\hat{c}_k$ is the predicted index of the beam (the index of the largest value in $\mathbf{p}$), $c^\star_k$ is the ground truth, and $\mathbb I$ denotes the indicator function. Note that the SNR of the mmWave downlink data transmission is set to be 0 dB through out the whole simulation, which may be different from the SNR of the uplink sub-6GHz signal or the mmWave pilots.

\subsection{Prediction Performance}

\begin{figure}[t]
\centerline{\includegraphics[width=12cm]{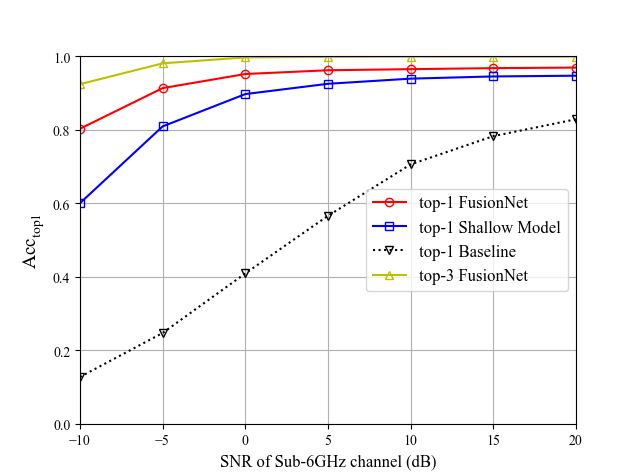}}
\captionsetup{font={small}}
\caption{\small The $Acc_{top1}$ predicted using FusionNet, shallow model and DNN proposed in \cite{alrabeiah2019deep}. The top-3 accuracy of FusionNet is also shown in this figure.}
\label{figcompare}
\end{figure}

Fig. \ref{figcompare} compares prediction accuracy of the FusionNet with other neural network architectures, where the top-3 accuracy of FusionNet is also displayed. The baseline curve is the performance of the DNN in  \cite{alrabeiah2019deep} that merely takes the sub-6GHz channel as input while the ``shallow model"  curve stands for predicting directly from the concatenated sub-6GHz and mmWave channel as shown in Fig. \ref{fig2}(a). Both the ``shallow model" network and the FusionNet are trained with $\tilde{N}_m=8$ active mmWave antennas and with  pilot SNR $= 20$dB. From Fig. \ref{figcompare}, the prediction accuracy of ``shallow model" network  and the FusionNet both outperform the baseline curve. Since the shallow model does not fully exploit the individual features of the two channels, the ``shallow model" curve is always below the FusionNet curve.  Moreover, the top-3 accuracy of FusionNet reaches 100$\%$ even when the SNR of the sub-6GHz channel is merely 0 dB, hence we will omit the top-3 accuracy in the rest of the simulation.

\begin{figure}[t]
\centering
\subfigure[$Acc_{top1}$]{
\includegraphics[width=10cm]{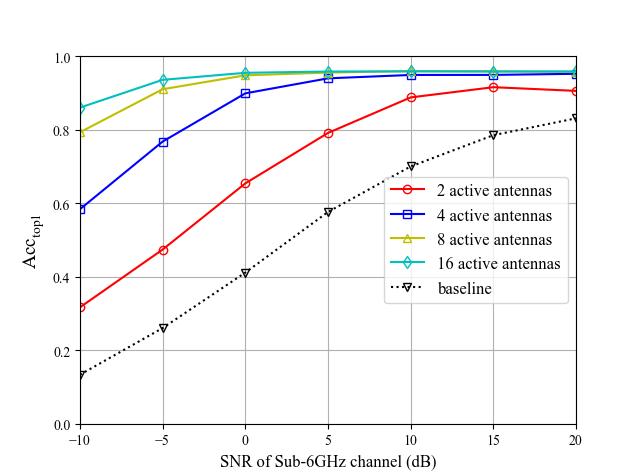}
}
\subfigure[Achievable rate]{
\includegraphics[width=10cm]{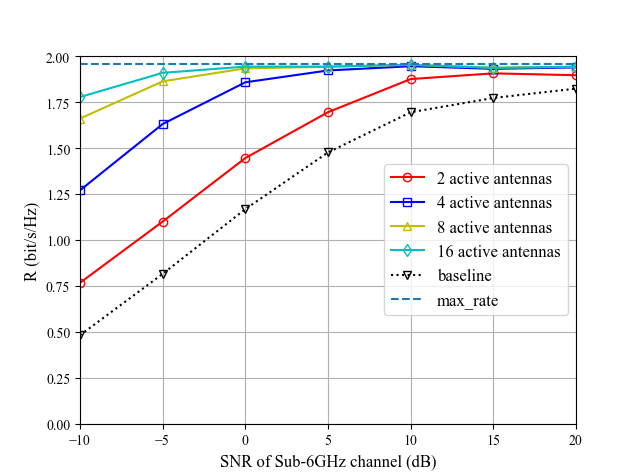}
}
\captionsetup{font={small}}
\caption{\small Prediction performance with different number of active mmWave antennas}
\label{fig4}
\end{figure}

Fig. \ref{fig4} displays prediction accuracy and the corresponding achievable rate versus SNR of the sub-6GHz channel estimation. The number of active antennas $\tilde{N}_m$  is 2, 4, 8, 16, respectively, and the mmWave pilot SNR is 20 dB. From Fig. \ref{fig4}(a), prediction accuracy of the FusionNet with any number of $\tilde{N}_m$  is always much better than the baseline method, especially at the low SNR region. An approximately 5 dB SNR gain in the sub-6GHz band is observed in terms of beam prediction accuracy even if we turn on only  2 mmWave antennas.  Moreover, the beam prediction accuracy of the FusionNet significantly improves as the number of active mmWave antennas increases while the improvement slows down beyond 8 active mmWave antennas. The achievable rate in Fig.~\ref{fig4}(b) follows the similar trend.  Another observation is that there is almost no rate loss in mmWave downlink data transmission when the SNR of the sub-6GHz signal is merely 5 dB and when $\tilde{N}_m=4$ mmWave antennas are used for beam calibration. All these observations clearly demonstrate the effectiveness of the proposed FusionNet.

\begin{figure}[t]
\centering
\subfigure[$Acc_{top1}$]{
\includegraphics[width=10cm]{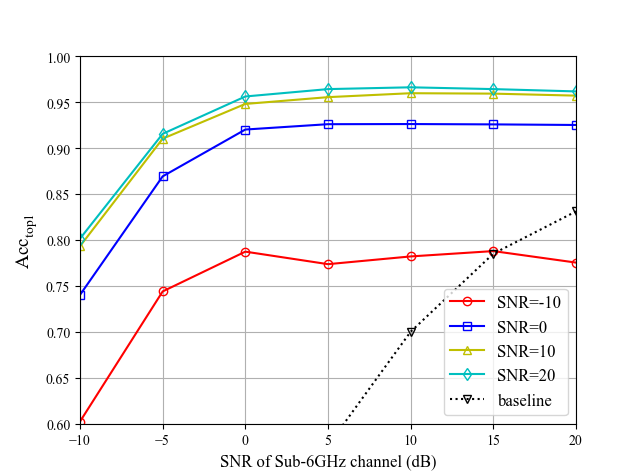}
}
\subfigure[Achievable rate]{
\includegraphics[width=10cm]{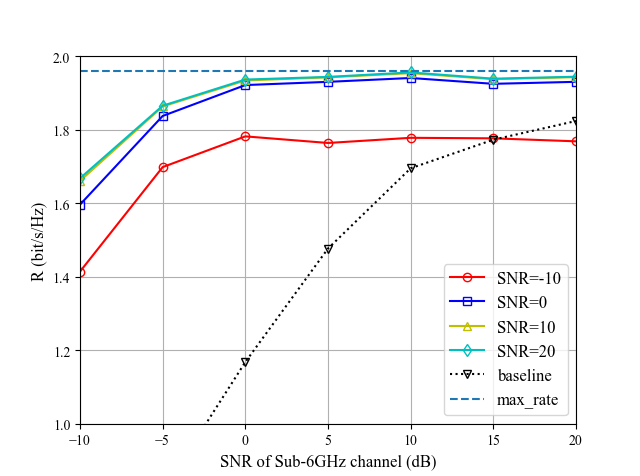}
}
\captionsetup{font={small}}
\caption{\small Prediction accuracy and achievable rate with the SNR of the mmWave channel}
\label{fig5}
\end{figure}

Fig.~\ref{fig5} depicts prediction accuracy and the achievable rate versus the sub-6GHz channel SNR  under different mmWave pilot SNR with 8 active mmWave antennas. From the figure, the FusionNet significantly outperforms the the baseline for most cases except when the mmWave pilot SNR is extremely low and the sub-6GHz training SNR is very high, which is not unexpected since in this case the calibration effect is not accurate enough, and then may drag down the performance predicted from the sub-6GHz channel. Nevertheless, the calibration effect with a low SNR is still positive when the sub-6GHz SNR is low. Therefore, the mmWave pilots would greatly help enhance the performance of a pure DNN in most practical scenarios.
Moreover, further increasing the pilot SNR beyond 5 dB does not present more positive effect, which is essential for mmWave transmission when there is a severe path loss.  Similar observations can also be found in the achievable rate in Fig.~\ref{fig5}(b).

\subsection{Effects of the Number of OFDM Pilot Subcarriers on Both Frequency Bands}
In previous examples, the FusionNet is examined with fully-loaded pilot subcarriers in both frequency bands. However, the practical protocol may assign limited number of pilot subcarriers in each OFDM block and it is of interest to check whether the FusionNet still works in this case.

\begin{table*}[t]
    \centering
    \captionsetup{font={small}}
    \caption{The prediction accuracy when using different number of mmWave subcarriers}
    \label{tablemmwsub}
    \begin{tabular}{ c  c  c  c  c  c  c  c }
\hline
\hline
      sub-6GHz SNR(dB) &  -10 & -5 & 0 & 5 & 10 & 15 &20\\
\hline
      $Acc_{top1}$ using all subcarries & 0.601 & 0.777 & 0.902 & 0.942 & 0.956 & 0.956 & 0.960 \\
\hline
      $Acc_{top1}$ using 1/8 subcarries & 0.563 & 0.761 & 0.896 & 0.944 & 0.952 & 0.956 & 0.956 \\
\hline
      $Acc_{top1}$ using 1/16 subcarries & 0.535 & 0.756 & 0.896 & 0.939 & 0.951 & 0.951 & 0.954 \\
\hline
      $Acc_{top1}$ using 1/32 subcarries & 0.529 & 0.744 & 0.883 & 0.924 & 0.937 & 0.939 & 0.940 \\
\hline
\hline
\end{tabular}

\end{table*}

Table. II demonstrates the FusionNet's performance when a fraction of OFDM subcarriers of the mmWave band are used as pilots, with $\tilde N_m=4$ active antennas. From the table, prediction accuracy drops very little compared to the fully loaded pilots even if we use only one pilot because that the beam calibration process mainly uses the angular information of the mmWave channel, thus the additional OFDM subcarriers cannot provide further prediction improvement. Therefore, in practice, one may reduce the mmWave pilot number to enhance the data throughput.

Table. III shows the FusionNet's performance using a fraction of OFDM subcarriers of the sub-6GHz channel. All OFDM subcarriers on $\tilde N_m=4$ mmWave active antennas  are used as pilots. Different from Table. II, prediction accuracy drops when the SNR of the sub-6GHz channel is low, however, it ceases as the sub-6GHz SNR increases. This is because the sub6-network of FusionNet might easily fit to the noise at the low SNR region, using more OFDM subcarriers for training will help reduce the effect of noise.

\begin{table*}[t]
    \centering
    \captionsetup{font={small}}
    \caption{The prediction accuracy when using different number of sub-6GHz subcarriers}
    \label{tablesub}
    \begin{tabular}{ c  c  c  c  c  c  c  c }
\hline
\hline
      sub-6GHz SNR(dB) &  -10 & -5 & 0 & 5 & 10 & 15 &20\\
\hline
      $Acc_{top1}$ using all subcarries & 0.601 & 0.777 & 0.902 & 0.942 & 0.956 & 0.956 & 0.960 \\
\hline
      $Acc_{top1}$ using 1/2 subcarries & 0.498 & 0.682 & 0.856 & 0.929 & 0.956 & 0.956 & 0.960 \\
\hline
      $Acc_{top1}$ using 1/4 subcarries & 0.424 & 0.589 & 0.780 & 0.898 & 0.949 & 0.956 & 0.960 \\
\hline
      $Acc_{top1}$ using 1/8 subcarries & 0.389 & 0.495 & 0.676 & 0.830 & 0.923 & 0.954 & 0.959 \\
\hline
\hline
\end{tabular}

\end{table*}

\subsection{ Utilizing Channel Sparsity and Data Augmentation}
In this part, two proposed data pre-processing approaches are evaluated.
The augmentation rate, $R_{aug}$, is defined as the size of the augmented dataset divided by the size of the original dataset.

\begin{figure}[t]
\centerline{\includegraphics[width=12cm]{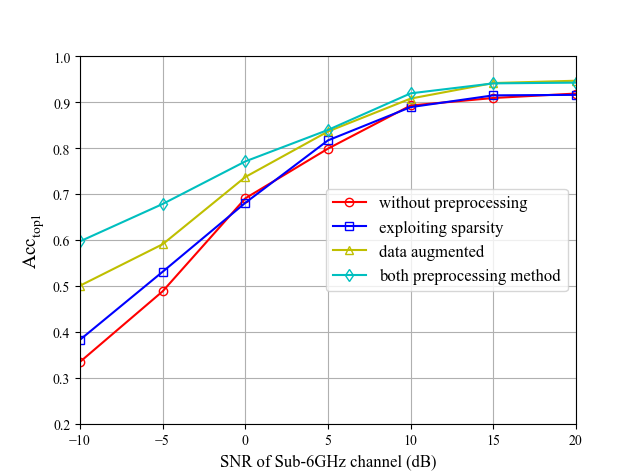}}
\captionsetup{font={small}}
\caption{\small  The prediction accuracy using sub-6GHz channel and the channel on 2 active mmWave antennas with different data pre-processing methods. }
\label{fig.prepro}
\end{figure}

Fig.~\ref{fig.prepro} shows the prediction performances utilizing channel sparsity, data augmentation, as well as the combination of these two approaches. For this simulation, ${\tilde N}_m=2$, the mmWave pilot SNR is 20 dB and the augmentation rate is 8. From the figure, the prediction performance is improved by exploiting the  channel sparsity and adopting data augmentation approach especially at the low sub-6GHz SNR, it can be further improved combining the two pre-processing approaches.

\begin{figure}[t]
\centerline{\includegraphics[width=12cm]{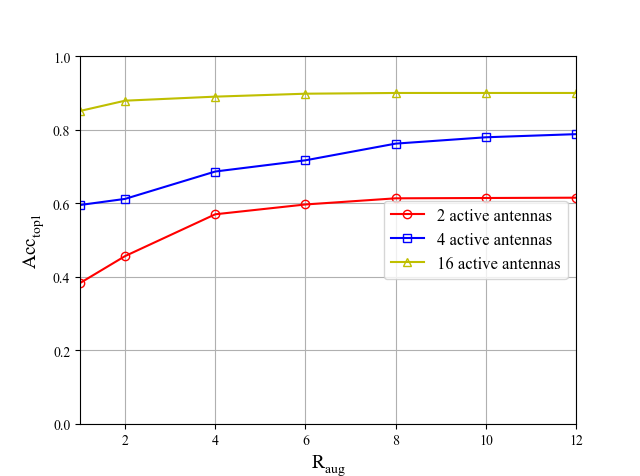}}
\captionsetup{font={small}}
\caption{\small  The prediction accuracy versus the data augmentation rate.}
\label{figaug}
\end{figure}

Fig.~\ref{figaug} displays prediction accuracy versus the augmentation rate when the SNR of the sub-6GHz and the mmWave channel are $-10$ dB and 20 dB, respectively, where the channel sparsity is also exploited. The prediction accuracy improves rapidly at first and then slows down with 2 or 4 active antennas, showing the synthesized data has fully exploited the underline information after certain rate. However, when $\tilde{N}_m$ increases to 16, the improvement brought by data augmentation is limited since the prediction accuracy  with the original dataset is good enough.

\subsection{Prediction Directly from Active MmWave Pilots}

\begin{figure}[t]
\centering
\subfigure{
\includegraphics[width=12cm]{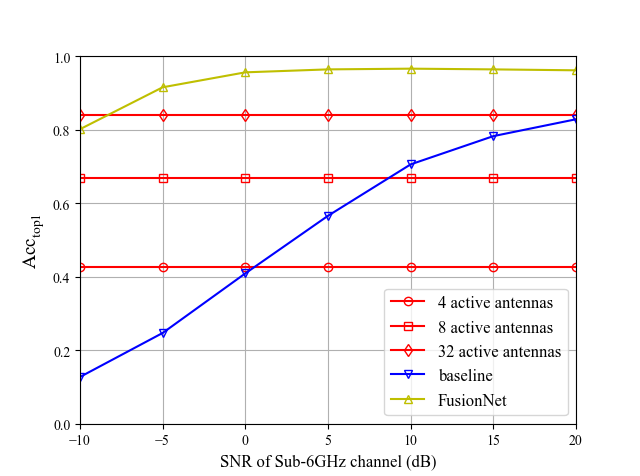}
}
\captionsetup{font={small}}
\caption{\small Prediction directly from mmWave channel}
\label{fig6}
\end{figure}

With the significant improvement brought by a few mmWave pilots, one natural question arises: is the enhancement because  $\tilde{\mathbf{h}}_m[k]$ itself is good enough to predict the optimal downlink beam? To answer this question, the prediction performance using mmWave channel only along with the baseline and the FusionNet, under pilot SNR 20 dB is shown in Fig.~\ref{fig6} where the FusionNet adopts 8 active mmWave antennas.

From the figure, the prediction accuracy based on the active mmWave antennas is not satisfactory, which is still below 85 $\%$ even if half of the total mmWave antennas (i.e., $\tilde{N}_m=32$) are active. In brief,  Fig.~\ref{fig6} clearly demonstrates the intriguingly novel aspect of the FusionNet, which can merge the two ``mediocre'' ways and results in an extremely precise prediction.

\section{Conclusion}
In this paper, we develop a deep learning based approach using the uplink sub-6GHz channel with very few pilots in the mmWave band to greatly enhance the performance of mmWave downlink beam prediction. Specifically, we design a novel DNN architecture, the FusionNet, that concatenates both the sub-6GHz and partial mmWave channel as the inputs. By extracting the individual features from the two different channels and perform concatenation,  the prediction accuracy and the achievable rate of the FusionNet outperforms all current state-of-art method. To further improve the prediction performance, we introduce a data augmentation approach to prevent over-fitting when training the FusionNet to extract and exploit the sparsity features of the channels. We show that even when the SNR of sub-6GHz and the mmWave channels are low, the proposed FusionNet is still able to predict the best beam using very few mmWave pilot with high fidelity, making itself a promising candidate for future full spectrum wireless applications.


\linespread{1.4}

\normalem
\bibliography{ref}

\end{document}